\documentclass[12pt]{article}

\usepackage{amssymb,amsmath,amsthm}
\usepackage[dvips]{graphics}
\usepackage[cp1251]{inputenc}
\usepackage[english]{babel}

\begin{document}

\author{Kostyakov~I.V.,~ Kuratov~V.V.}
\title{Massive Yang-Mills fields, translational  and nonsemisimple gauge symmetry}

\maketitle

\begin{abstract}
Gauge fields of semisimple groups of internal symmetries are massless and require the
special techniques for guarantee their mass. Massive mechanisms usually contain
transfor\-mati\-ons of shifts typical to nonsemisimple groups. We show that under
the localization of nonsemisimple internal symmetry the gauge fields corresponding to
translation generators  are massive. In addition, we introduce nonlinear
generalizations of well-known models, with local translational symmetry and as a result,
the massive gauge fields. Thus, the local Galilean symmetry is realized on a special pair
of scalar fields, leading to massive electrodynamics, and the localization of the
Euclidean group leads to massive non-Abelian theory without matter fields. We propose a
simple interpretation of the Stueckelberg mechanism. 
\end{abstract}

\section{ Introduction}
First we describe the basic concepts and properties of gauge theories \cite{SlavnovFaddeev}.
We turn our attention to the notion of mass and motivate the study of translational and nonsemisimple gauge theories.

\subsection{Gauge fields} 
 The principle of local gauge invariance is a fundamental dynamic principle of modern
physics and allows to describe the fundamental interactions of nature. Let a matter field
$ \varphi(x) $ is the $N$-vector which transforms under the action of the group $ G $ of
matrix $ N\times N $ of internal symmetries
\begin{equation}
\varphi(x)\rightarrow g\varphi(x)=e^{T_k\alpha_k}\varphi(x),
\label{1}
\end{equation}
where $ T_k $ are the  generators of group $ G $, and $ \alpha_k $ are the group parameters. Derivative
$ \displaystyle{\partial=\frac{\partial}{\partial x^\mu}} $ similarly transformed 
\begin{equation}
\partial\varphi \rightarrow\partial(e^{T_k\alpha_k}\varphi(x))=e^{T_k\alpha_k}\partial\varphi(x).
\label{2}
\end{equation}
 Now let $\alpha_k=\alpha_k(x): $
\begin{equation}
\varphi(x)\rightarrow g(x)\varphi(x)=e^{T_k\alpha_k(x)}\varphi(x),
\label{3}
\end{equation}
but
\begin{equation}
\partial\varphi(x)\rightarrow\partial(g(x)\varphi(x))=g(x)\partial\varphi(x)+\varphi(x)\partial g(x)
\end{equation}
and for the  `` right '' transformation of the derivative it is necessary to extend one
by introducing the vector gauge Yang - Mills field \ \
$ A = A_\mu^k T_k $ \cite{YangMills} with the value in the Lie algebra of $ G $ and transformed by the formula
\begin{equation}
 A\rightarrow gA g^{-1}- \partial g\cdot g^{-1}.
\label{4}
\end{equation}
Then the covariant derivative transformed multiplicatively
\begin{equation}
\nabla  \varphi=( \partial - A) \varphi \rightarrow g\nabla  \varphi,
\label{5}
\end{equation}
curvature or tensor  $F$ of Yang - Mills fields is determined by
\begin{equation}
F \varphi=[\nabla,\nabla] \varphi = [\partial-A,\partial-A] \varphi,
\label{7b}
\end{equation}
and its Lagrangian is
\begin{equation}
L_{YM}=-\frac{1}{4}Tr\, F^2.
\label{7}
\end{equation}
Describing the matter fields, gauge fields, and their interaction theory is given by Lagrangian
\begin{equation}
L=\frac{1}{2}\nabla  \varphi^*\nabla  \varphi+V( \varphi)-\frac{1}{4}TrF^2 .
\label{7a}
\end{equation}
If $ G = SO (2) \approx U (1) $  is a group of two-dimensional rotation, then (\ref{7b}) is electromagnetic tensor, and
(\ref{7}) is a Lagrangian of the electromagnetic field. The group $ SU (2) \times U (1) $ correspond to the standard
model of unified electroweak interaction.

Mass term
$
\frac{m^2}{2}A^2
$
is not invariant under transformation (\ref {4}), therefore gauge field is massless. Of the four known interactions,
electromagnetism and gravity are the long-range, and hence massless. For the weak and strong interactions it is
necessary to seek ways of origination mass.

\subsection{ The mass and nonsemisimple group}
Free field equations and kinematics of the fields are determined by their transformation properties under the Poincare
transformations  $ ISO (3,1) $ of the motion group of Minkowski space-time. It is well known that the representations of
the simple group $ SO (5) $ are described by two "spins" $ (j_1, j_2) $ with integer and half-integer values.
Corresponding to elementary particles fields  transforms under representations of the nonsemisimple Poincare group  $
ISO (3,1) $ defined by a spin and a continuous parameter $ M $ indentified with mass  $ (j, M). $ Transition to
nonsemisimple group or contraction $ SO (5) \rightarrow ISO (3,1) $ may be thought as induces by transition $
(j_1, j_2) \rightarrow (j, M) $ and generates a mass of particles from the second spin at the kinematic level.

It turns out that such  phenomenon there is on a dynamic level. If interactions are introduced by the gauge invariance
principle, the transition from a simple gauge group to nonsemisimple one containing translation transformation  can also
generate a mass of gauge field. Massive fields corresponding to translational transformations are deal with in
\cite{Nelipa} in constructing of Lagrangian of gauge fields for nonsemisimple gauge groups. In our case, a mass term
originates from  the variables describing the interaction of gauge field with a field of matter.

Consider the local translational transformation 
\begin{equation}
\varphi(x)\rightarrow  \varphi(x)+\alpha(x), 
\label{a}
\end{equation}
then the covariant derivative is defined as
\begin{equation}
\nabla \varphi=\partial \varphi-A,
\end{equation}
transformation of gauge field
\begin{equation}
A\rightarrow A+\partial\alpha,
\label{b}
\end{equation}
and Lagrangian has the form
\begin{equation}
L=\frac{1}{2}(\partial \varphi-A)^2.
\label{c}
\end{equation} 
 Quadratic in $A$ term can be interpreted as a mass one.
 Note that not only the kinematical group of $ ISO (3,1) $, but 
Higgs and Stueckelberg mechanism  also contain global or local transformation of shift important to generate a
mass.
  Nonsemisimple symmetry rarely has been the subject of investigation, but some of their unusual properties are increasingly attracting attention.
  Features of nonsemisimple gauge theories discussed in the papers \cite{Tseytlin, Witten, Nuyts, Ferrari}.
  In \cite{Ferrara} the relationship of spontaneous symmetry breaking  and the transition from semisimple to
nonsemisimple symmetry is studied.

In addition teleparallel gravity is progressing actively \cite{Pereira}.
  This theory is in fact the gauge theory of shift. Note the work \cite{Scaria} where the relationship between 
Stueckelberg mechanism, gauge symmetry, and translational transformations have been investigated. Nonsemisimple
symmetries also arise in the compactification of the like Kaluza--Klein theories   \cite{Candelas}.
  
\section{ Models}
In this section we present the well-known globally and locally invariant theory on the example of  Abelian $U(1)$ and the Galilean $G(1)$ transformation and non-abelian $O(3)$
model written in radial coordinates suitable for the study the translation mass mechanisms. Parameterized by function $
V(\rho) $ Lagrangians are introduced and they are nonlinear generalizations of the known theories.

\subsection{  Global theories } 
 Global $ U (1) $ symmetry is realized by the phase transformation of the comp\-lex-va\-lued field $ \psi $ or
the rotation
in the space $ (\varphi_1, \varphi_2) $ of pairs of scalar fields
\begin{equation}
\psi= \varphi_1+i \varphi_2,
\label{9}
\end{equation}
\begin{equation}
\psi\rightarrow e^{i\alpha}\psi=\left\{\begin{array}{l}
 \varphi_1\rightarrow  \varphi_1\cos\alpha + \varphi_2\sin\alpha \\
 \varphi_2\rightarrow  \varphi_1\sin\alpha- \varphi_2\cos\alpha, \end{array} \right.
\label{10}
\end{equation}
with Lagrangian
\begin{equation}
L=\frac{1}{2} \partial\psi^* \partial\psi=\frac{1}{2}( \partial \varphi_1)^2+\frac{1}{2}( \partial \varphi_2)^2.
\label{11}
\end{equation}
Now write down all the formulas in the radial variables
\begin{equation}
\psi=\rho e^{i\varphi},
\label{12}
\end{equation}
\begin{equation}
\psi\rightarrow e^{i\alpha}\psi=\left\{\begin{array}{l}
\rho \rightarrow\rho \\
\varphi\rightarrow \varphi +\alpha, \end{array} \right. 
\label{13}
\end{equation}
\begin{equation}
L=\frac{1}{2}( \partial\rho)^2+\frac{1}{2}\rho^2( \partial\varphi)^2.
\label{14}
\end{equation}
Note that the most general $U(1)$-invariant Lagrangian has the form
\begin{equation}
L=\frac{1}{2} \partial\rho^2+\frac{1}{2}V(\rho)\rho^2( \partial\varphi)^2-U(\rho),
\label{15}
\end{equation}
and corresponds to non-Euclidean metric in the space of fields $ (\varphi_1, \varphi_2). $

Global $\Gamma(1)$ symmetry is realized as the phase transformation or as rotation  in the space of $ (\varphi_1, \varphi_2) $  which now appear to be Galilean transformation by replacing a complex unit  $ i^2 =- 1 $ on the dual one
\begin{equation}
\psi= \varphi_1+\iota \varphi_2,
\label{16}
\end{equation}
\begin{equation}
\psi\rightarrow e^{\iota\alpha}\psi=\left\{\begin{array}{l}
 \varphi_1\rightarrow  \varphi_1 \\
 \varphi_2\rightarrow  \varphi_2+\alpha \varphi_1.\end{array} \right.
\label{17}
\end{equation}
Choose a Galilean invariant Lagrangian in the form of
\begin{equation}
L=\frac{1}{2}( \partial \varphi_1)^2+\frac{1}{2}V( \varphi_1) \varphi_1^2\left( \partial\left(\frac{ \varphi_2}{ \varphi_1}\right)\right)^2-U( \varphi_1).
\label{18}
\end{equation}
 If we move again in the radial coordinates
\begin{equation}
\psi=\rho e^{\iota\varphi}=\rho +\iota\rho\varphi,
\label{19}
\end{equation}
then after replacement
\begin{equation}
 \varphi_1=\rho,\quad\quad \frac{ \varphi_2}{ \varphi_1}=\varphi,
\label{20}
\end{equation}
we get the transformation of translations
\begin{equation}
\psi\rightarrow e^{\iota\alpha}\psi=\left\{\begin{array}{l}
\rho\rightarrow \rho \\
\varphi\rightarrow  \varphi +\alpha, \end{array} \right.
\label{21}
\end{equation}
do not change the Lagrangian
\begin{equation}
L=\frac{1}{2}( \partial\rho)^2+\frac{1}{2}V(\rho)\rho^2( \partial\varphi)^2-U(\rho).
\label{22}
\end{equation}
Similarity of the formulas for both groups is not accidental and the transformation (\ref{20}) can be considered as a
substitute for the variables that are $ U(1)$ in $ \Gamma(1) $ and vice versa. The only difference of groups - $U(1)$
is a compact and its parameter changes within the
  $ 0 \leq \alpha \leq 2 \pi $ while in the group $\Gamma(1)$  $ -\infty \leq \alpha \leq +\infty. $

 $O(3)$ invariant theory is given by Lagrangian
 \begin{equation}
L=\frac{1}{2}\left( \partial\vec{\phi}\cdot \partial\vec{\phi}\right)=\frac{1}{2}\left( \partial{\phi}_i\cdot \partial{\phi}_i\right),
\label{22a}
\end{equation}
where the vector  
 $ \vec{\phi}^t= \left(\phi_1,\phi_2,\phi_3 \right)  $
 transforms by the ortogonal matrix
 $g$, $\phi\rightarrow g \phi$,
 $g^tg=1$. 
 Let us pass to the radial variable in the isospace 
 $\vec{\phi}=\rho\vec{\sigma}$.
If $\vec{e}_3^{\, t}=(0,0,1)$ and a unit vector of field  
$\vec{\sigma}$ 
connects with an element of rotation group 
$\Omega=\Omega(\theta,\varphi)$ by equality  $\vec{\sigma}=\Omega\vec{e}_3$, 
then
$$
\partial\phi=\Omega\left(\partial\rho+\rho\Omega^{-1}\partial\Omega\right)=\Omega\left(\partial\rho+\rho\Lambda\right),
$$
here
$$
\Lambda=\Omega^{-1}\partial\Omega.
$$
In radial coordinates the Lagrangian  has the form
$$
L=\frac{1}{2}\left(\partial\rho\right)^2+\frac{1}{2}\rho^2\Lambda^2.
$$
The generalized Lagrangian
$$
L=\frac{1}{2}\left(\partial\rho\right)^2+\frac{1}{2}V(\rho)\rho^2\Lambda^2.
$$
 also will have $O(3)$ invariance.

\subsection{ Local theories }
Localization of both $U(1), \Gamma(1)$ Lagrangians with a  choice $V(\rho^2)=1$ gives scalar electrodynamics
\begin{equation}
\psi \rightarrow e^{i\alpha(x)}\psi=\left\{\begin{array}{l}
\rho\rightarrow \rho \\
\varphi\rightarrow  \varphi +\alpha(x), \end{array} \right.
\label{23}
\end{equation}
\begin{equation}
 \partial\psi\rightarrow D\psi=( \partial - A)\psi=\left\{\begin{array}{l}
 \partial\rho\rightarrow  \partial\rho \\
 \partial\varphi\rightarrow  D\varphi= \partial\varphi - A, \end{array} \right.
\label{24}
\end{equation}
\begin{equation}
 A\rightarrow  A+ \partial\alpha,
\label{25}
\end{equation}
\begin{equation}
L=\frac{1}{2}\left(\partial\rho\right)^2+\frac{1}{2}\rho^2( \partial\varphi
- A)^2-\frac{1}{4}F^2.
\label{26}
\end{equation}
The theory is compact in the $U(1)$ case and noncompact for the $\Gamma(1)$ group \cite{Polyakov}.
In radial variables both models are the gauge theories of localized transformations of shifts of matter fields.
Let $B=A-\partial\phi$, then $F^A=[\partial-A,\partial-A]=F^B=[\partial-B,\partial-B]$ and 
we have a theory of charged scalar field interacting with electromagnetic field
\begin{equation}
L=\frac{1}{2}\left( \partial\rho\right)^2+\frac{1}{2}\rho^2 B^2-\frac{1}{4}F^2.
\label{26a}
\end{equation}

In Cartesian coordinates a local $O(3)$ theory looks like
$$
L=\frac{1}{2}(\nabla\vec{\varphi}\nabla\vec{\varphi})+L_{YM},
$$
where $\nabla=\partial -A, \quad A=A^\alpha_\mu T^\alpha$, and $T^\alpha $ are the generators of algebra $O(3).$

In radial coordinates the generalized Lagrangian looks 
\begin{equation}
L=\frac{1}{2}(\nabla\varphi)^t(\nabla\varphi)=\frac{1}{2}\left( \partial\rho\right)^2+\frac{1}{2}V(\rho)\rho^2 \left(B_2^2+B_3^2\right)-\frac{1}{4}F^2.
\end{equation}

\section{A mass}
All known mass mechanisms are quite simple but  some additional structures are used. New scalar
fields (Stueckelberg) are introduced,  handpicked potential of self-action  are added to these fields (Higgs) or some
restrictions are imposed  (Gromov). We briefly explain them.

\subsection{  Stueckelberg mechanism }
70 years ago Stueckelberg \cite{Stueckelberg} noticed that if the scalar field $\varphi $ that interacts with the
4-potential is added to the Lagrangian of electromagnetic fields $A$
\begin{equation}
L=-\frac{1}{4}F^2+\frac{1}{2}m^2( \partial\varphi - A)^2,
\label{27}
\end{equation}
then under the transformations
\begin{equation}
 A\rightarrow  A+ \partial\alpha(x),
\label{28}
\end{equation}
\begin{equation}
\varphi\rightarrow \varphi +\alpha(x),
\label{29}
\end{equation}
the gauge invariance are preserved, i.e.  in theory there is a localized shift symmetry.
Then you can choose the gauging $\varphi=0$ or you can go to a new variable $ B= A- \partial\varphi$,
after which we get
\begin{equation}
L=-\frac{1}{4}F^2+\frac{1}{2}m^2 B^2
\label{30}
\end{equation}
where $B$ is a massive electromagnetic field. 
Gauge field $ B $ is transformed identically
$ B \rightarrow  B$. 
The field $ \varphi $ has disappeared, providing a mass for field $B$.
The formulas (\ref{28})--(\ref{30}) are identical to (\ref{a})--(\ref{c}), so the Stueckelberg mechanism can
be interpreted as the theory of the local translational symmetry of the scalar field.

\subsection{Higgs mechanism }
We dwell on the elementary abelian version of the mechanism. Details and the non-Abelian ge\-ne\-ra\-li\-za\-ti\-ons
can be
found in \cite{Higgs}.
Consider $U(1)$ locally invariant theory in radial coordinates
\begin{equation}
L=\frac{1}{2}(\nabla \psi)^2-\frac{F^2}{4}-U(|\psi|)
= \frac{1}{2}( \partial\rho)^2+\frac{\rho^2}{2}( \partial\varphi - A)^2-\frac{F^2}{4}-U(\rho),
\label{31}
\end{equation}
and let there is a minimum in the self-action at  $\rho=\rho_0$
\begin{equation}
\min U(\rho)=U(\rho_0).
\label{32}
\end{equation}
Then the fluctuation field $\rho$ near the vacuum value  $ \rho_0 $ can be considered.
Going to the variable $ R $ by the transformation of shift 
\begin{equation}
R=\rho-\rho_0
\label{33}
\end{equation}
 and introducing the field $B$
\begin{equation}
 B= A- \partial\varphi
\label{34}
\end{equation}
we have a Lagrangian
\begin{equation}
L=\frac{1}{2}( \partial R)^2+\frac{1}{2}(R+\rho_0)^2{ B}^2-U(R+\rho_0)-\frac{1}{4}F^2
\label{35}
\end{equation}
with quadratic mass term $\frac{1}{2}\rho^2_0 B^2$.
Note that $F^A=F^B$.

\subsection{Gromov mechanism}
Recently \cite{Gromov}  more simple way of the origination of mass of the gauge field was proposed.
If in the formula (\ref{31}) we assume
\begin{equation}
\rho=\mbox{ const }=m,
\label{36}
\end{equation}
which is equivalent to the imposition of constraint, we have
\begin{equation}
L=\frac{m^2}{2}( \partial\varphi - A)^2-\frac{1}{4}F^2
\label{37}
\end{equation}
and after the mentioned above replacement 
\begin{equation}
 B= A- \partial\varphi,
\label{38}
\end{equation}
we obtain
\begin{equation}
L=\frac{m^2}{2}B^2-\frac{1}{4}F^2.
\label{39}
\end{equation}
	
Higgs mechanism is considered the most reliable, due to similarities in solid state physics, but it is unclear the
origin of the minimum of potential. 
In addition, the particle of the field $\varphi $, Higgs boson has not yet been discovered.
Stueckelberg and Gromov mechanisms formally look the same but they have different interpretations.

It should be noted the work  \cite{Faddeev}, which offered an unusual approach to the problem of
masses of gauge fields.

\section{ New mass mechanisms}
In this section two ways of the origination of mass are described.
Mass terms in Lagrangians can be obtained both  a special choice of functions $V(\rho)$ in the generalized
models described earlier and the direct  construction of nonsemisimple gauge theories.

\subsection{Massive electrodynamics}

The second term in (\ref{26a}) describes the interaction of matter fields $\rho$ with the gauge field $B$.
However, this term can be turned into a mass one.
Take a Lagrangian (\ref{22}) with $V(\rho)\rho^2=1$.
Note that this does not impose on the $\rho$ no conditions, just a different choice of functions $ F $ correspond to
various Galilean invariant Lagrangian
\begin{equation}
L=\frac{1}{2}( \partial\rho)^2+\frac{1}{2}( \partial\varphi - A)^2-\frac{1}{4}F^2.
\end{equation}
Gauging $\varphi=0$ or the replacement (\ref{34})  gives
\begin{equation}
L=\frac{1}{2}( \partial\rho)^2+\frac{1}{2} A^2-\frac{1}{4}F^2.
\end{equation}
The field $\rho$ does not interact with the $A$, therefore the derived massive electromagnetic field has only a
imaginary interaction with $\varphi$ and this field can be removed  by gauge transformation.
Comparing the formulas, we see that a Shtueckelberg mechanism can be interpreted also as the construction
of a gauge field with Galilean Group.
In this case it is massive.
The field $\rho$, which in the case of the Higgs mechanism interacts with
$A$, and therefore must be observed, is now free, and hence unobserved. Based on the $U(1)$  local
Lagrangian (\ref{15}), the choice  $V(\rho)\rho^2=1$ corresponds to non-Euclidean metric in the space of
fields.

\subsection{Massive $O(3)$ model}
Consider $O(3)$ locally invariant model.
Let us take a radial representation for $\vec{\phi}$ as in the global case.
Let $\vec{e}_3^{\, t}=(0,0,1)$, and the unit vector of field $\vec{\sigma}$ is related with
element of rotation group $\Omega=\Omega(\theta,\varphi)$ by the
equality  $\vec{\sigma}=\Omega\vec{e}_3 $.
 Then 
$$
\vec{\phi}=\left(\begin{array}{c}
\phi_1\\
\phi_2\\
\phi_3 \end{array}\right) = \rho\vec{\sigma}=\rho\Omega\vec{e}_3=
\rho\left( \begin{array}{ccc}
\cos\varphi & \sin\varphi & 0 \\
-\sin\varphi & \cos\varphi & 0 \\
0 & 0 & 1 \end{array}\right)\times
$$
$$
\times\left( \begin{array}{ccc}
1 & 0 & 0 \\
0&\cos\theta & \sin\theta  \\
0&-\sin\theta & \cos\theta  \end{array}\right)\cdot
\left( \begin{array}{ccc}
\cos\alpha & \sin\alpha & 0 \\
-\sin\alpha & \cos\alpha & 0 \\
0 & 0 & 1 \end{array}\right)
\cdot\left(\begin{array}{c}
0\\
0\\
1 \end{array}\right) .
$$
Covariant derivative is given to the species
\begin{equation}
D\vec{\phi}=\left(\partial+ A\right)\rho\Omega\vec{e}_3=
\Omega\left(\partial\rho+\rho\left(\Omega^{-1}\partial\Omega+
\Omega^{-1} A\Omega\right)\right)\vec{e}_3.
\end{equation}
Denote by
$$
 B=\Omega^{-1} \partial\Omega+
\Omega^{-1} A\Omega,
$$
write down the Lagrangian
$$
L=\frac{1}{2}(D\vec{\phi})^tD\vec{\phi}=\frac{1}{2}
\left( \partial\rho\right)^2+\frac{1}{2}\rho^2\left(B^2_2+B^2_3\right)+L_{YM}.
$$
Generalization usually looks like
$$
L=\frac{1}{2}
\left( \partial\rho\right)^2+\frac{1}{2}V(\rho)\rho^2\left(B^2_2+B^2_3\right)+L_{YM},
$$
and it allows under the special choice of $V(\rho)\rho^2=m^2$ to obtain massive gauge fields. Here $L_{YM}$ is
the standard Lagrangian (\ref{7}).

\subsection{Euclidean gauge field}
Consider now the non-Abelian nonsemisimple Euclidean gauge group.
The assumption of non\-semi\-simple nature of the group of internal symmetries was studied in \cite{Sogami}. Let us add
to the rotation transformation (\ref{10}) the shifts
$$
\psi \rightarrow e^{i\alpha}\psi +t=\left\{\begin{array}{l}
\phi_1\cos\alpha +\phi_2\sin\alpha +t_1 \\
-\phi_1\sin\alpha +\phi_2\cos\alpha +t_2.
\end{array} \right.
$$
We get  Euclidean group $E(2)$, motion group of Euclidean metric, and the group of invariance of Lagrangian
(\ref{11}). Localization of the symmetry leads to the two massive fields, $A_1, \, A_2$,
  which corresponds to the shifts and rotations massless field $ A_0 $.
  Global $E(2)$ transformation
$$
\left(\begin{array}{c}
\phi_1\\
\phi_2\\
1 \end{array}\right) \rightarrow 
\left( \begin{array}{ccc}
\cos\alpha & \sin\alpha & t_1 \\
-\sin\alpha & \cos\alpha & t_2 \\
0 & 0 & 1 \end{array}\right)\left(\begin{array}{c}
\phi_1\\
\phi_2\\
1 \end{array}\right) 
$$
with generators
$$
T_0=\left(\begin{array}{ccc}
0 & 1 & 0 \\
-1 & 0 & 0 \\
0 & 0 & 0 \end{array} \right), \quad
T_1=\left(\begin{array}{ccc}
0 & 0 & 1 \\
0 & 0 & 0 \\
0 & 0 & 0 \end{array} \right), \quad
T_2=\left(\begin{array}{ccc}
0 & 0 & 0 \\
0 & 0 & 1 \\
0 & 0 & 0 \end{array} \right),
$$
and derivative
$$
\partial\Phi=
\left(\begin{array}{ccc}
\partial & 0 & 0 \\
0 & \partial & 0 \\
0 & 0 & \partial \end{array} \right)
\left(\begin{array}{c}
\phi_1\\
\phi_2\\
1 \end{array}\right) \rightarrow \partial (g\Phi)=g\partial \Phi,
$$
defines the simplest type of $E(2)$ globally invariant Lagrangian
$$
L=\frac{1}{2}\left(\partial\Phi\right)^t\left(\partial\Phi\right).
$$
Local variant normally requires the extension of the derivative, taking into account the type of group generator
$$
\nabla\Phi=
\left(\begin{array}{ccc}
\partial & -A_{0} & -A_{1} \\
A_{0} & \partial & -A_{2} \\
0 & 0 & \partial \end{array} \right)
\left(\begin{array}{c}
\phi_1\\
\phi_2\\
1 \end{array}\right) \rightarrow 
$$
$$
\rightarrow 
\nabla' (g\Phi)=(\partial+A')g\Phi=
(\partial g +A'g +g\partial)\Phi=
$$
$$=
g(g^{-1}\partial g +g^{-1}A'g +\partial)\Phi=g\nabla \Phi.
$$
whence it follows  the transformation (\ref{4}) for the gauge fields.
The Lagrangian is
$$
L=\frac{1}{2}\left(\nabla\Phi\right)^t\left(\nabla\Phi\right)=
$$
\begin{equation}
=
\frac{1}{2}\left(\partial\phi_1 -A_0\phi_2+A_1\right)^2+\frac{1}{2}\left(\partial\phi_2 +A_0\phi_1+A_2\right)^2+\tilde{L}_{YM}(F^A).
\end{equation}
The presence of quadratic terms with respect to $A$ indicates about massiveness "translati\-onal" compo\-nent of
gauge fields. Here $\tilde{L}_{YM}$ is no longer determined by the formula (\ref{7}).

Put  another formulas for the transformation of fields $A$:
\begin{eqnarray*}
	A_0 & \rightarrow &A_0+\partial\alpha,\\
	A_1 & \rightarrow & A_1\cos\alpha+A_2\sin\alpha-A_0t_2-t_2\partial\alpha+\partial t_1, \\
  A_2 & \rightarrow & -A_1\sin\alpha+A_2\cos\alpha+A_0t_1+t_1\partial\alpha+\partial t_2.
\end{eqnarray*}
If we choose  new variables using a gauge transformation $\alpha = 0$,
$t_1=\phi_1$, $t_2=\phi_2$
\begin{eqnarray*}
	B_1 & = & A_1-A_0\phi_2+\partial \phi_1, \\
  B_2 & = & A_2+A_0\phi_1+\partial \phi_2,
\end{eqnarray*}
we have the Lagrangian
\begin{equation}
L=\frac{1}{2}\left(B_1^2+B_2^2\right)+\tilde{L}_{YM}(F^B),
\end{equation}
where the two degrees of freedom $\phi_1, \phi_2$ passes into the transverse components of
fields  $B_1, B_2$, giving them a mass.

\section{Conclusions}
We propose massive generalization of gauge theories with semisimple symmetry group. Localization of nonsemisimple
internal symmetries also lead to the masses of translational component of gauge fields.
In both cases does not require the introduction of additional fields.
The possibility of using  described mechanisms in realistic models requires additional research. Separate task is to
write the nonsemisimple Lagrangians for Yang-Mills fields due to lack of nondegenerate bilinear Killing form on 
the nonsemisimple algebras.

\vspace{5mm}

The work has been partially supported by RFBR grant 08-01-90010 --- Belarus and the
program '' Mathematical problems of nonlinear dynamics`` of the Presidium of Russian
Academy of Sciences.


\begin{thebibliography}{99}

\bibitem{SlavnovFaddeev}
Fadeev L.D., Slavnov A.A. Gauge Fields: An Introduction to Quantum Theory. 1994 
  
\bibitem{YangMills}
   Yang~C.N. and  Mills~R.L.
Conservation of Isotopic Spin and Isotopic Gauge Invariance.
Physical Review.   N96. 1954.  Pp. 191-195.

\bibitem{Nelipa}  Demichev~A.P., Nelipa~N.F.
Methods of constraction of gauge-invariant lagrangians for arbitrary Lie groups. 
Progress of Theoretical Physics. V.76, N3. 1986. Pp.715-725.


\bibitem{Tseytlin}  Tseytlin~A.A. On gauge theories for non-semisimple groups.
Nuclear Physics B. V.450. N~1-2. 1995. Pp.~231-250.

\bibitem{Witten}  Nappi C.R., Witten E.
Wess-Zumino-Witten model based on a non-semisimple group.
Physical Review Letters. V.71. 1993. Pp.~3751-3753.

\bibitem{Nuyts}  Nuyts J., Wu T.T.
Yang-Mills theory for nonsemisimple group.
Physical Review D. V.67. N~2. 02014. 2003. 9~p.

\bibitem{Ferrari}  Ferrari F. 
Chern-Simons field theories with nonsemisimple gauge group
of symmetry.
Journal of Mathematical Physics. V.44. N1. 2003. Pp.~138-145.

\bibitem{Ferrara}  Andrianopoli L., Ferrara S., Lledo M.A., Macio O.
Integration of massive states as contraction of nonlinear $\sigma$ models. 
Journal of Mathematical Physics. V.46. 072307.  2005. 33~p.

\bibitem{Pereira} Andrade~V.C., Guillen~L.C.T. and~Pereira~J.G.
 Teleparallel gravity: an
overview. ArXiv:gr-qc/0011087.

\bibitem{Scaria} Scaria~T. Translational groups of gauge transformations. Physical Review D. V.68. 105013. 2003.
12p. ArXiv:hep-th/0302130. 


\bibitem{Polyakov} 
Polyakov~A.M.   Gauge Fields and Strings. Harwood Academic Publishers, 1987. 


\bibitem{Stueckelberg}  Stueckelberg E.C.
Helv. Phys. Acta. V.11. 1938. Pp.226,229.

\bibitem{Higgs}  Higgs~P.W. 
Broken symmetry and the masses of gauge bosons. 
Physical Review Letters. V.13. 1964. p.508. \hspace{2mm} 
 Englert, Brout R. \hspace{2mm} Broken symmetry and the masses of gauge vector mesons.//
Physical Review Letters. V.13. 1964. p.321.  \hspace{2mm}
Peskin~M.E.,  Schroeder~D.V.  An Introduction To Quantum Field Theory.

\bibitem{Gromov}  Gromov~N.A. ArXiv:0705.4575 [hep-th], arXiv:hep-th/0611092, hep-th/0611079.


\bibitem{Faddeev}  Faddeev~L.D. An alternative interpretation of the Weinberg--Salam model.
ArXiv:0811.3311 [hep-th]. \hspace{2mm}  Chernodub~M.N., Faddeev~L.D., Niemi~A.J.  
Non-Abelian Supercurrents and de Sitter Ground State in Electroweak Theory. //ArXiv:0804.1544 [hep-th]. 


\bibitem{Sogami}  Sogami~I.S A non-semisimple hidden symmetry for flavor physics. Progress of theoretical physics.
V.114. N4. 2005. Pp. 873--887.

\bibitem{Candelas}  Candelas P.,  Perevalov E.,  Pajesh G. hep-th/9703148.

\end{thebibliography}
\end{document}